\documentclass[Final]{eps}

\usepackage{graphicx}
\usepackage{times}

\newcommand       \erg          {\,{\rm erg}}

\newcommand       \s            {\,{\rm s}}

\newcommand       \mum          {\,{\rm \mu m}}

\newcommand       \simali       {\sim\,}

\newcommand       \kabs         {\kappa_{\rm abs}}

\newcommand       \md           {m_{\rm d}}
\newcommand       \Td           {T_{\rm d}}
\newcommand   \xstar        {{\small \left[{\rm X/H}\right]}}
\newcommand       \nx           {n_{\rm\small X}}
\newcommand       \mH           {m_{\rm\small H}}
\newcommand       \mHstar       {M_{\rm\small H}}
\newcommand       \mud          {\mu_{\rm d}}
\newcommand       \Etot         {E_{\rm tot}}

\volumenumber{60}
\publishyear{2008}
\frompage{\pageref{firstpage}}
\topage{\pageref{finalpage}}

\shorttitle{Jiang, Zhang, \& Li:
            The 21$\mum$ and 30$\mum$ circumstellar dust features}

\title{The 21$\mum$ and 30$\mum$ circumstellar dust features
       in evolved C-rich objects}
\author{B.W. Jiang$^1$, Ke Zhang$^1$, and Aigen Li$^2$}
\affiliation{$^1$Department of Astronomy, Beijing Normal University,
                 Beijing 100875, China\\
             $^2$Department of Physics \& Astronomy,
                 University of Missouri,
                 Columbia, MO 65211, USA}
\received{xxxx xx, 2008}\revised{xxxx xx, 2008}\accepted{xxxx xx, 2008}
\abstract{%
The 21$\mum$ and 30$\mum$ bands are the strongest dust emission
features detected in evolved low- and intermediate-mass C-rich stars
(i.e. asymptotic giant branch [AGB] stars, proto-planetary nebulae
[PPN], and planetary nebulae [PN]). While the 21$\mum$ feature is
rare and exists only in the transient PPN phase, the 30$\mum$
feature is more common and seen in the entire late stage of stellar
evolution, from AGB to PPN and PN phases, as well as in the
low-metallicity galaxies: the Large Magellanic Cloud (LMC) and the
Small Magellanic Cloud (SMC). The carriers of these features remain
unidentified. Eleven of the twelve well-identified 21$\mum$ sources
also emit in the 30$\mum$ band, suggesting that their
carriers may be somewhat related. 
}

\begin{document}
\label{firstpage} \maketitle \copyrighttext{}

\section{Introduction}
The circumstellar chemistry of the low- and intermediate-mass stars
from AGB on is dichotomized into O-rich and C-rich classes according
to the C/O abundance ratio, since C and O would combine at
relatively high temperature to form the very stable molecule CO. The
dominant dust species in O-rich circumstellar envelope (CSE) is
silicate, abundant in both amorphous and crystalline forms. In
C-rich objects, the C-bearing molecules (e.g. C$_{2}$, C$_{3}$, CN,
and some hydrocarbon compounds) dominate, while the most common dust
feature should be assigned to the SiC 11.3$\mum$ band detected in
more than 700 stars most of which are C-rich objects. However, the
strongest and most enigmatic spectral features of C-rich objects are
the 21$\mum$ and 30$\mum$ emission bands. The combined contribution
of these two bands can amount to more than 20\% of the total
infrared flux, so they greatly influence the CSE evolution.
Furthermore, the dispersion of the CSE dust to the interstellar
medium (ISM) brings the dust for these features enter into the
recycling of cosmic dust.

\section{The 21$\mum$ feature}
The 21$\mum$ feature was first identified in 4 PPNe from the
IRAS/LRS spectra in 1989 (Kwok et al.\ 1989). Later, from the high
resolution spectra taken by ISO/SWS, this feature is accurately
measured to have a peak wavelength at $\simali$20.1$\mum$. But the
tradition to name it as the 21$\mum$ feature is retained. As one of
the rarest dust features, so far, this feature has been well
identified only in a dozen sources (see Kwok et al.\ 1999).
Consistent with its rare appearance, the 21$\mum$ sources all are at
a short phase in stellar evolution, i.e. PPN, which lasts only
several $\simali$10$^3$ years. They show a double-peaked spectral
energy distribution (SED), with one optical peak arising from the
radiation of the central star (with an effective temperature of
$\sim$\,5500--7500\,K) and the other (usually mid-) infrared (IR)
peak from the low-temperature dust envelope. Besides, all these
21$\mum$ PPNe are C-rich, metal-poor but rich in s-process elements,
a quite small subgroup of PPNe (Kwok et al.\ 1999).

Since its discovery, the 21$\mum$ feature has attracted much
attention for its unidentified carrier. The number of proposed
carrier candidates exceeds the number of the feature emitters (i.e.
a dozen). The proposed candidates fall into three broad categories:
(i) organic dust, (ii) oxides, and (iii) the silicon and carbon
compound.
Because the 21$\mum$ emission band is found only in C-rich sources,
the C-bearing organic dust species are naturally considered as the
carrier candidates, including hydrogenated fullerenes, polycyclic
aromatic hydrocarbon (PAH), hydrogenated amorphous carbon (HAC),
synthetic carbonaceous macromolecules, and amides (especially urea).
In addition, inorganic candidates are proposed: diamonds, iron
oxides ($\gamma$-Fe$_{2}$O$_{3}$, Fe$_{3}$O$_{4}$, FeO), SiS$_{2}$,
titanium carbide nanoclusters, doped SiC, and SiC core-SiO$_2$
mantle particles, some of which also contain carbon atoms. Posch et
al. (2004) discussed in details many of these proposed candidates.
The proposal for these materials to be the 21$\mum$ feature
candidate is generally based on their experimental spectra which
show a feature around 20$\mum$ similar to the observed 21$\mum$
feature profile. But a valid carrier needs not only to be able to
reproduce the observed feature, but also (1) to be abundant enough
to account for the observed intensity and (2) to not have any
associated features, consistent with the observational fact the
21$\mum$ band has no appreciable associated features.
For example, since titanium is a rare element in the universe,
Ti-bearing dust species are often examined in terms of the required
abundance (relative to H), e.g., titanium carbide (TiC) nanoclusters
were rejected because the cosmic Ti abundance is not enough to
account for the intensity of the observed 21$\mum$ emission feature
(see Li 2003, Hony et al.\ 2003, Chigai et al.\ 2003).

\subsection{Carrier}
We have investigated the feasibility of all the inorganic carrier
candidates based on their elemental abundance requirement and the
possible presence of any associated features (see Zhang et al.\ 2008
for details).
The mass of the carrier, $m_{\rm d}$, which is directly related to
the abundance of the elements composing the carrier, is the key
parameter to determine the emitted power of the feature through
\begin{equation}\label{eq_Etot}
    E_{\rm tot} = m_{\rm d}\int_{\rm 21\mum\,band}
    \kabs(\lambda)\times4\pi B_{\lambda}(\Td)d\lambda
\end{equation}
where $E_{\rm tot}$ is the total emitted power of the 21$\mum$
feature, $\kabs(\lambda)$ is the mass absorption coefficient of the
proposed dust carrier, $B_{\lambda}(\Td)$ is the Planck function at
wavelength $\lambda$ and dust temperature $\Td$ (here we assume that
the 21$\mum$ feature-emitting dust has a single temperature of
$\Td$). The abundance of element $X$ locked up in dust $\xstar$ to
account for the observed power is then
\begin{equation}\label{eq_abundance}
  \xstar\equiv\frac{\nx\,\md/\mud \mH}{\mHstar/\mH} = \frac{\nx\,\Etot} {\mud\,\mHstar
  \int_{21\mum\,{\rm band}} \kabs(\lambda)\times
  4\pi B_{\lambda}(\Td) \,d\lambda}
\end{equation}
where $n_{\rm X}$ is the number of $X$ atoms per molecule, $\mud$ is
the molecular weight of the dust, $M_{\rm H}$ is the total H mass of
the circumstellar envelope of the star.

Therefore, with the mass absorption coefficient $\kabs$
experimentally measured or theoretically calculated, the required
abundance of specific element $X$ can be calculated from the
observed feature intensity. For the strongest 21$\mum$ feature
source HD\,56126 (also known as IRAS\,07134+1005), the total power
emitted through the 21$\mum$ band is $\Etot \approx
10^{36}\erg\s^{-1}$ (i.e., about 300\,L$_\odot$). From this
abundance requirement, both the TiC nanoclusters and the fullerenes
coordinated with Ti atoms originally suggested by Kimura et al.
(2005b) are disqualified to be the carrier due to the shortage of
element Ti; the same fate is for SiS$_{2}$ from the shortage of
element S (SiS$_{2}$ also can not satisfy the criterion on the
associated features) (Zhang et al.\ 2008).

Although the 21$\mum$ feature is discovered only in C-rich sources
that exhibit additional spectral features arising from C-bearing
molecules, it has no clear association with any other features. As
we will discuss later, ten among the twelve known 21$\mum$ sources
emit at the 30$\mum$ feature as well, but their intensities are not
correlated at all. On the other hand, the proposed carrier
candidates often exhibit additional spectral feature(s) (aside from
the 21$\mum$ feature). A typical example is the SiC dust
(responsible for the most popular dust feature in C-rich sources at
11.3$\mum$). The suggestion for SiC to be the 21$\mum$ feature
carrier must pass the examination that the expected intensity ratio
of the 11.3$\mum$ feature to the 21$\mum$ feature should agree with
the observed value of $I(11.3\mum)/I(21\mum)$\,$<$\,0.01 (Jiang et
al. 2005).

Detailed calculations have shown that this is impossible from the
optical properties of SiC (Jiang et al.\ 2005). The same examination
has been applied to other inorganic carrier candidates. This
requirement rules out all the candidates which exhibit associated
feature(s), e.g., SiS$_{2}$ with an associated feature at
16.8$\mum$, SiC dust with carbon impurities at 11.3$\mum$, carbon
and silicon mixtures at 9.5$\mum$, SiC core-SiO$_{2}$ mantle grains
at 8.3$\mum$ and 11.3$\mum$, Fe$_{2}$O$_{3}$ at 9.2$\mum$, 18$\mum$
and 27.5$\mum$, Fe$_{3}$O$_{4}$ at 16.5$\mum$  and 24$\mum$ are all
rejected as a viable candidate (see Zhang et al.\ 2008 for details).
This lack of agreement is evident since the 21$\mum$ feature is
usually weak compared to the associated features in the experimental
spectra of the proposed candidate materials while in the observation
it is the opposite.

With the two criteria (i.e. the requirement of abundance to account
for the observed high intensity of the 21$\mum$ feature and the
requirement of no clear associated feature) taken into account,
among the inorganic carrier candidates only FeO (as originally
proposed by Posch et al.\ 2004) survives, as both Fe and O are rich
elements and FeO has no associated feature. However, it is not clear
how FeO can form and survive in a C-rich environment where all O
atoms are consumed to form CO. The
 summary of the examination results on the inorganic carrier
candidates is described in Zhang et al.\ (2008).

The same examination procedure should be performed for the organic
carrier candidates. From the abundance point of view, the organic
materials are not expected to have problem since they are composed
mainly of carbon and hydrogen, both of which are abundant elements.
But they may have a big problem due to displaying associated
features that are not observed in the 21$\mum$ sources. For example,
PAHs, as one of the organic carrier candidates, show a distinctive
series of features at 3.3, 6.2, 7.7, 8.6 and 11.3$\mum$. In fact,
ten among the twelve 21$\mum$ sources have such PAH features (mostly
around 3.3, 7.7, and 11.3$\mum$), which makes it reasonable to list
PAHs as a carrier candidate. But the intensity ratios of the
21$\mum$ feature to the prominent features at 3.3--11.3$\mum$ of
PAHs can be problematic. Other organic species may have similar
dilemma.

\subsection{Search for new emitters}

 Although the twelve well-accepted 21$\mum$ emitters are all PPNe,
 a few PNe and extreme AGB stars have been reported to have this
 feature as well.
 Hony et al.\ (2001) argued that the 21$\mum$
 feature is seen in two PNe. Volk et al.\ (2000) reported the
 detection of this feature in two highly evolved carbon stars.
 It is therefore necessary to search new 21$\mum$ emitters
 in both PPNe and objects in other late stages of stellar evolution.
 To clarify the stellar evolutionary stages
for this feature to exist, which discloses the physical and chemical
conditions of the carrier, would be very helpful in the
identification of the carrier.

 We used the \emph{Spitzer} Space Telescope to search for new 21$\mum$
 emitters. The adopted criteria which could potentially
detect new 21$\mum$ emitters was to find the
 sources with an emission excess around 21$\mum$ which may
 come from the 21$\mum$ feature. The mid-IR space project Midcourse
 Space Experiment (MSX) provided an opportunity to choose such sources
 because one of its broad photometric bands, band E, has a central
 wavelength of 21.34$\mum$ and a bandwidth of 6.24$\mum$
 that covers the entire wavelength
range of the 21$\mum$ feature. So the task of picking out the
objects with 21$\mum$ emission excess
 was transferred to searching for the sources with excess in the MSX-E band.
 To be more explicit, the observed objects were selected
 based on the following procedure:

\begin{enumerate}
\item Pick up the evolved stars from the IRAS PSC ({\it Point
      Source Catalog}). Since the 21$\mum$ sources are not
      necessarily limited to the post-AGB phase,
      stars in a broad evolutionary stage,
      from evolved AGB stars to PNe,
      are selected based on the IRAS color indices;
      specifically, $\log F_{25}/F_{12}$\,$>$\,$-0.2$
      and $\log F_{60}/F_{25}$\,$<$\,0.
      The combination of these two color indices are to exclude
      YSOs and galaxies and leave with us only evolved stars
      (van der Veen \& Habing 1988).
      The selected stars should have high-quality IRAS photometry
      at 12$\mum$, 25$\mum$, and 60$\mum$.
\item Look for the counterparts of these sources in the MSX PSC
      catalog. The cross-association is done by position-matching.
      Only those sources in the sky area surveyed by MSX,
      mostly in the Galactic plane, are left.
      The selected sources should have high quality MSX photometry
      in the MSX C (12.13$\mum$) and E (21.34$\mum$) bands.
\item Find the sources with excess in the MSX E band.
      Fig.\,1 is the color-color diagram of
      $\log F_{\rm 25}/F_{\rm 12}$ vs.
      $\log F_{\rm msxE}/F_{\rm msxC}$ of the sources chosen above.
      We fit these data by a quadratic line
      which approximately represent the mean values
      of the $\log F_{\rm msxE}/F_{\rm msxC}$ color.
      Apparently, the objects lying above this line
      are more likely to have the 21$\mum$ emission feature.
      As the MSX C band (12.13$\mum$) is an analog to
      the IRAS 12$\mum$ band, and the dust continuum
      radiation in the MSX E band (21.34$\mum$) should not differ
      much from that in the IRAS 25$\mum$ band,
      the objects lying above the fitting line are considered to
      have an excess emission in the MSX E band.
\item Choose the sources whose colors are similar to those of
      the known 21$\mum$ sources.
      Because most of the known 21$\mum$ sources are at high latitudes,
      they were not observed by MSX.
      We therefore calculate $\log F_{\rm msxE}/F_{\rm msxC}$
      for the known 21$\mum$ sources by convolving their ISO/SWS
      spectra with the MSX Relative Spectral Response function.
      It can be seen that, as expected, these known 21$\mum$ sources
      (denoted by asterisks) are lying mostly above or on
      (depending on the strength of the feature) the fitting line.
      In comparison, the C-rich PPN IRAS\,19477+2401 without 21$\mum$
      emission (denoted by a triangle) is well below the fitting line.
      In addition, the color indices $\log F_{\rm 25}/F_{\rm 12}$
      of the known 21$\mum$ sources
      are all larger than 0.5 (because the dust is cold).
      Therefore, the sources with $\log F_{\rm 25}/F_{\rm 12}$
      between 0.5 and 1.0 are chosen.
\item Exclude the sources which are O-rich or other early-type
      objects according to the SIMBAD database.
      A few of the above-selected sources, which are previously
      observed and found to be OH/IR stars or Wolf-Rayet objects, are rejected.
\item Exclude the sources brighter than 100\,Jy in the MSX E band.
      This is to avoid the saturation problem in the IRS high
      resolution mode.
\end{enumerate}

\begin{figure*}[t]
\centerline{\includegraphics[width=14.5cm,clip]{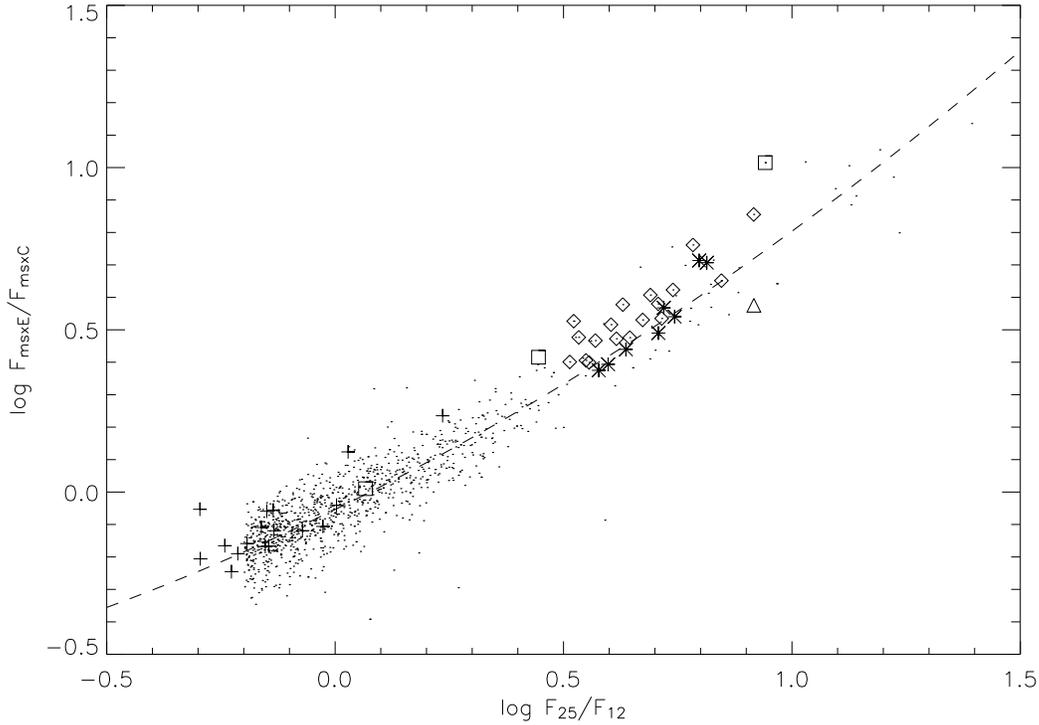}}
\caption{Color-color diagram of the sources with good quality
         measurements both in the IRAS 12, 25$\mum$ bands,
         and in the MSX C (12.13$\mum$) and E (21.34$\mum$) bands,
         with the dashed line being a quadratic fit to these data.
         The selected candidates are labelled by diamond-surrounded dots.
         Others symbols -- asterisk: known strong 21$\mum$ sources;
         cross: AGB stars with 10 and 20$\mum$ silicate emission;
         square: O-rich PPNe with silicate emission;
         triangle: C-rich PPN without 21$\mum$ emission.}
\end{figure*}

As the result, we were left with 18 sources which are possibly
evolved AGB stars, PPNe and PNe and have excess emission around
21$\mum$. They are shown in Fig.\,1 as diamond-surrounded dots. Two
of the 18 sources were listed in others' {\it Spitzer} GO (General
Observer) proposals with higher priority. The other 16 sources were
observed from our {\it Spitzer} proposal (ID P--30403). They were
observed using the Infrared Spectrograph (IRS) with two modules:
Short-High (SH) and Long-High (LH) during Cycle-3 from October 2006
to May 2007. The data were taken in ``staring mode'' with two nod
positions along each IRS slit. Due to the anticipated high
brightness of the objects, no specific off-set observations were
dedicated. Our observations found no new objects showing the
21$\mum$ emission feature, while some objects have features
attributed to crystalline silicates which bring about at least part
of the MSX-E band excess (Jiang et al.\ 2008). It's worthy noting
that Hrivnak et al.\ (2008) reported two new detections of the
21$\mum$ feature based on their {\it Spitzer} observations. Although
we are not yet at a position to conclude that the 21$\mum$ feature
exists only in C-rich PPNe, it at least suggests that this feature
appears only under rigorous conditions and the sources are really
rare.

In summary, we conclude that the carrier candidates suggested up to
now are mostly unacceptable and the conditions for the existence of
the 21$\mum$ feature are not very clear either. To make progress in
the study of this feature, more laboratory experiments are needed to
search for materials that can meet the abundance and spectral
requirements. In addition, sophisticated dust model may be helpful.
From the observational side, larger scale surveys should be
performed to the objects sharing the characteristics of the 21$\mum$
sources, i.e. metal-poor,  the s-process elements enhanced, C-rich,
PPN and so on. But the \emph{Spitzer} spectral surveys of the
evolved stars in LMC and SMC did not find new emitters. Since the
evolved stars in LMC and SMC are mainly C-rich and very likely
metal-poor, this result may indicate that these two characteristics
are not the key or not enough. Perhaps a survey of the C-rich PPN
would be a nice project to discover new 21$\mum$ emitters.

\section{The 30$\mum$ feature}
The 30$\mum$ feature is not as rare as the 21$\mum$ feature. Since
its discovery (Low et al.\ 1973) from ground-based observations,
this feature has been detected in 63 Galactic objects: 36 evolved
carbon stars, 14 PPNe and 13 PNe (Hony et al.\ 2002), thanks to {\it
KAO} (Kuiper Airborne Observatory), {\it ISO} (Infrared Space
Observatory) and {\it Spitzer}. Moreover, the 30$\mum$ emission
feature is also found in evolved stars of other galaxies, including
17 carbon-rich stars in the LMC (Zijlstra et al.\ 2006) and 8
carbon-rich stars in the SMC (Sloan et al.\ 2006; Lagadec et al.\
2007).

The 30$\mum$ feature shows up not only in the PPN phase, but also in
the AGB and PN phases. The conditions for its presence is not as
strict as the 21$\mum$ feature. However, like the 21$\mum$ feature,
it shows up only in C-rich environment. Meanwhile, they do not form
a special sub-group of the evolved C-rich objects, at least partly
due to the lack of a detailed analysis of the 30$\mum$ sources.
Nevertheless, there seems to be a borderline of dust temperature for
the 30$\mum$ emission. In our Galaxy, this feature occurs from
extreme AGB stars on to later stages. In LMC and SMC, only in the
coolest stars was this feature detected. Given the uncertainties in
determining the specific dust temperature, the highest dust
temperature of the 30$\mum$ emitters should not exceed 700\,K. The
temperature upper limit means the feature carrier exists only in low
temperature.

The spectral profile, both the peak and width, of the 30$\mum$
feature changes in different objects. The feature peak shifts to
longer wavelengths with lower dust temperature for the Galactic
sources. The shape changes significantly from object to object that
it even causes debates about the feature profile. One suggestion is
that the feature is composed of two sub-features at different
wavelengths according to the {\it ISO} observation of some PPNe
(Volk et al.\ 2002); the other suggestion is that the feature is
monolithic but varies with the temperature and shape of the dust
carrier (Hony et al.\ 2002). The experimental absorption spectra of
MgS, a promising carrier candidate, show great dependence on dust
shape (Kimura et al.\ 2005a).

MgS seems to be a widely accepted carrier of the 30$\mum$ feature,
although other candidates have also been proposed (including organic
molecules naturally present in C-rich environment). MgS was first
brought forward by Goebel \& Moseley (1985) based on its IR
resonance band. Later observations of the 30$\mum$ band were often
explained in terms of the MgS dust in the radiative transfer
modelling, but the modelling results bring about a few questions on
MgS as the carrier.

\begin{description}
\item [Shape] Spherical-shaped dust can not fit the observations well,
      while CDE (continuous distribution of ellipsoids) can,
      because the optical properties of MgS change greatly with the dust shape.
      However, the shape of the MgS dust formed in laboratory is
      often roundish, network-like or cubic other than
      ellipsoidal (Kimura et al.\ 2005a).
      Moreover, the CDE shape distribution
      is difficult to explain the observed sub-structures
      (but see Zhukovska \& Gail 2008).
      It thus leads to the suspicion of the reality of
      the ellipsoidal shape assumption.
\item [Abundance] For the sources in the Galaxy,
      the required abundances of Mg and S to account for
      the observed intensity of the 30$\mum$ feature is
      usually much less than cosmic abundances of these elements.
      For example, Jiang et al.\ (1999) found that 10\% of
      the cosmic S abundance is sufficient.
      Thus the MgS abundance seems to have no problem.
      Nevertheless, the detection of the 30$\mum$ feature
      in LMC and SMC needs to re-consider the abundance issue
      since these two galaxies are much more metal-poor than
      our Galaxy, in particular the metallicity of
      the SMC is only about 10\% of our Galaxy.
      No modelling efforts have been taken for the 30$\mum$ features
      detected in the Magellanic clouds (MCs) partly due to
      the lack of abundance measurements of these sources.
      Quantitative modelling will be very helpful.
      Besides, whether sulfur is mainly tied to MgS is unknown.
      From the examination of the Wild\,2 dust collected by
      {\it Stardust}, most of the sulfur in the anhydrous dust
      is concentrated in iron sulfides while no MgS dust was found\textbf{;}
      besides, some of the anhydrous silicates are Mg-rich (Flynn 2008).
      This suggests that both Mg and S may be tied to
      dust species other than MgS. However, Zhukovska \& Gail (2008)
      pointed out that FeS is abundant only in O-rich environments; while in
      C-rich environment, MgS is more stable than FeS, and Mg should neither
      be tied to silicates formed in O-rich environment. This may alleviate the
      problem of abundance caused by consuming Mg and S to form other molecules.

\item [Associated features] The experimental IR spectra of MgS
      exhibit some sub-structures although the associated features
      depend on dust shape (Kimura et al.\ 2005a). This might be
      a piece of good news for the argument that the 30$\mum$ feature
      is composed of two sub-features, but whether the observed and
      experimental sub-structures agree with each other needs careful comparison.
\end{description}


\section{Comparison of the 21$\mum$ and 30$\mum$ features}
The 21$\mum$ and 30$\mum$ features are the strongest dust features
in evolved C-rich sources. They share some characters such as to
influence the final evolution of stars. But they are different in
several aspects as mentioned in previous sections. Here a detailed
comparison is summarized.

\paragraph{Profile}
The 21$\mum$ feature is not symmetrical, but has a steep rise in
blue wing and a long tail at the red end $\simali$23--24$\mum$. It
is certainly a one-component feature. More importantly, the profile
of this feature is universal, i.e. there is no clear change from
source to source. This leads one to believe that the carrier should
be a single sort of dust material with universal properties in
various sources. On the other hand, the profile of the 30$\mum$
feature changes so greatly from source to source that a
two-component structure was proposed as discussed above. The change
of the profile could be caused either by the dust properties such as
the shape and size distributions that lead to the change of optical
properties, or by the physical and chemical conditions such as the
temperature and metallicity. From the analysis of the 30$\mum$
sources in our Galaxy, Hony et al.\ (2002) found that the feature
profile is related to the dust temperature. In the low-metallicity
SMC, this feature appears to be weaker than in our Galaxy (Sloan et
al.\ 2006), but there is no significant difference in LMC and in our
Galaxy (Zijlstra et al.\ 2006) even though LMC is clearly metal poor
compared to our Galaxy. Therefore, it seems that the metallicity is
not directly correlated with the strength of the 30$\mum$ feature.

\paragraph{Emitters}
First of all, both features are detected only in C-rich sources,
and in the very late stages of evolution.
The 21$\mum$ feature is detected only in PPNe,
the very short transient phase of stellar evolution,
while the 30$\mum$ feature shows up from AGB to PPN and PN phases.
In addition, the 30$\mum$ feature is discovered in LMC and SMC
where the metallicity is low.
Meanwhile there has been no report on the discovery of
the 21$\mum$ feature in the MCs
although the 21$\mum$ sources in our Galaxy are metal-poor,
but this may be caused by the observation bias:
so far only evolved AGB stars (not PPNe) have been observed
in the MCs. It implies that the carrier of the 21$\mum$ feature
can survive only in a narrow range of physical conditions.
On the contrary, the dust for the 30$\mum$ feature can live
in various environments, in particular, its survival to
the PN phase would bring it to the ISM to participate
in the cosmic dust re-cycling.

\begin{table}[h] 
\renewcommand{\arraystretch}{1.2}
\vspace{-0.3cm}
\caption{The 21$\mum$ feature sources.
         The numbers under the ``30$\mum$''
         and ``21$\mum$'' labels are the fractions
         of the total IR flux emitted in
         the 30$\mum$ feature or
         the 21$\mum$ feature
         (in case no numbers are given,
          $\surd$: detected, ...: not yet detected).
         }
\vspace{-0.1cm}
\begin{center}
\begin{tabular}{cccc} \hline
  Object & 30$\mum$ & 21$\mum$ & Ref.  \\ \hline
  02229+6208 & 20 & 1 & Hrivnak et al. 2000\\
  04296+3429 & 22 & 5 & Szczerba et al. 1999 \\
  05113+1347 & ...  & $\surd$ & Kwok 1999 \\
  07134+1005 & 12 & 8 & Hrivnak et al. 2000 \\
  16594-4656 & 22 & 7 & Hrivnak et al. 2000 \\
  19500-1709 & 24 & 1 & Volk et al. 2002 \\
  20000+3239 & 19 & 1 & Hrivnak et al. 2000 \\
  AFGL 2688 & $\surd$ & $\surd$ & Omont et al. 1999 \\
  22223+4327 & $\surd$ & $\surd$ & Volk et al. 2002 \\
  22272+5435 & 24 & 3 & Szczerba et al. 1999 \\
  22574+6609 & 19 & 2 & Hrivnak et al. 2000 \\
  23304+6147 & 24 & 5 & Volk et al. 2002 \\
  \hline
\end{tabular}
\end{center}
\end{table}

\paragraph{Co-existence}
Among the twelve 21$\mum$ feature sources, ten sources show the
30$\mum$ emission feature. Early in 1995, the 16--46$\mum$ spectra
were taken for five 21$\mum$ emitters by {\it KAO}. It was found
that the 30$\mum$ feature is present in all of them, while the
intensity of these two bands are not correlated (Omont et al.\
1995). Later, from the ISO/SWS spectra, Hrivnak et al.\ (2000)
discovered new 30$\mum$ emitters and quantitatively analyzed the
percentage of the total IR flux emitted in the two features (see
Table 1). Up to now, 11 of the twelve 21$\mum$ emitters are reported
to have the 30$\mum$ emission feature. The only source that was not
classified as with the 30$\mum$ feature is IRAS\,05113+1347 which
was not observed in this spectral range. Therefore, the 11/12 ratio
is a lower limit of the percentage of the 30$\mum$ feature in the
21$\mum$ sources and a 12/12 ratio is possible. Such an almost 100
percentage that the 30$\mum$ feature appears in the 21$\mum$
emitters may lead to the proposition that their origins are related.
But this is challenged by the fact that their strengths are not
correlated. The most promising candidate carrier of the 30$\mum$
feature MgS is neither suggested as that of the 21$\mum$ feature as
it does not exhibit any features around 21$\mum$. Nevertheless, the
carriers for both features are not well identified. They may
(partly) share some organic molecules as the carrier, or their
carrier may co-exist.

\acknowledgments{BWJ and KZ are supported in part by China's grants
2007CB815406 and  NSFC 10473003. BWJ thanks the LOC for financial
support. AL is supported in part by NASA/\emph{Spitzer} Theory
Programs and NSF grant AST 07-07866. We thank Drs. B. Hrivnak and Y.
Kimura for helpful suggestions.}


\email{B.W. Jiang (e-mail: bjiang@bnu.edu.cn), Ke Zhang (e-mail:
zhangkebnu@gmail.com), and A. Li (e-mail: lia@missouri.edu)}
\label{finalpage}
\end{document}